\documentclass[%
 reprint,
superscriptaddress,
 amsmath,amssymb,
 aps,
]{revtex4-1}

\usepackage{lipsum}  
\usepackage{graphicx}
\usepackage{dcolumn}
\usepackage{bm}

\usepackage{hyperref}
\hypersetup{colorlinks=true, citecolor=blue, urlcolor=blue, linkcolor=blue}
\newcommand{\cabb}{Cs$_2$AgBiBr$_6$}

\begin{document}

\preprint{APS/123-QED}

\title{Anharmonicity and Ultra-Low Thermal Conductivity in Lead-Free Halide Double Perovskites}

\author{Johan Klarbring}
 \email{johan.klarbring@liu.se}
 \affiliation{%
Theoretical Physics Division, \\
Department of Physics, Chemistry and Biology (IFM),
Link\"{o}ping University, SE-581 83, Link\"{o}ping, Sweden
}%
\author{Olle Hellman}
 \affiliation{%
Theoretical Physics Division, \\
Department of Physics, Chemistry and Biology (IFM),
Link\"{o}ping University, SE-581 83, Link\"{o}ping, Sweden
}%
\author{Igor A. Abrikosov}
\affiliation{%
Theoretical Physics Division, \\
Department of Physics, Chemistry and Biology (IFM),
Link\"{o}ping University, SE-581 83, Link\"{o}ping, Sweden
}%
\affiliation{Materials Modeling and Development Laboratory, \\ NUST “MISIS”, 119049 Moscow, Russia}
\author{Sergei I. Simak}
 \affiliation{%
Theoretical Physics Division, \\
Department of Physics, Chemistry and Biology (IFM),
Link\"{o}ping University, SE-581 83, Link\"{o}ping, Sweden
}%

\date{\today}

\begin{abstract}
The lead-free halide double perovskite class of materials offers a promising venue for resolving issues related to toxicity of Pb and long-term stability of the lead-containing halide perovskites.
We present a first-principles study of the lattice vibrations in \cabb, the prototypical compound in this class, and show that the lattice dynamics of \cabb\ is highly anharmonic, largely in regards to tilting of AgBr$_6$ and BiBr$_6$ octahedra. Using an energy and temperature dependent phonon spectral function, we then show how the experimentally observed cubic-to-tetragonal phase transformation is caused by the collapse of a soft phonon branch. We finally reveal that the softness and anharmonicity of \cabb\ yield an ultra-low thermal conductivity, unexpected of high symmetry cubic structures. 

\end{abstract}

\maketitle
%
%
Lead halide perovskites, which have emerged as one of the most promising and intensively studied class of semiconductors for optoelectronic applications\cite{Huang2017,niu2015,Green2014}, are extremely anharmonic materials\cite{yang2017,Carignano2017,Yaffe2017,Klarbring2019,Bechtel2018}. The atomistic dynamics in these APbX$_3$ systems, where A is either Cs or an organic cation such as methylammonium (MA) or formamidinium (FA) and X $=$ F, Cl, Br or I, is highly complex, with large observed differences between local and average structures \cite{Bertolotti2017,beecher2016,Cottingham2018,Yaffe2017}. These systems typically have low frequency, often unstable, octahedral tilting modes and complex anharmonic movements of the A site ion, in particular related to the rotational dynamics of the organic molecule \cite{Yaffe2017}. This complex atomic motion has a large impact on electronic properties through a strong electron-phonon interaction \cite{Wright2016,Schlipf2018,Motta2015} and thus becomes highly influential on the optoelectronic performance of these materials.

The so-called lead free halide double perovskites (HDP) \cite{Slavney2016,Volonakis2016}, where the divalent Pb$^{2+}$ cations are replaced by pairs of mono- and trivalent cations such as Ag$^{+}$, Bi$^{3+}$, and In$^{3+}$, are envisioned to solve crucial issues of the single lead halide perovskites relating to poor long-term stability and the presence of toxic Pb. Unfortunately, they typically have bandgaps that are larger-than-optimal for photovoltaic applications \cite{Filip2016}. There is, however, a vast chemical space within this class of materials which can be explored in attempts to tune crucial properties. The electron-phonon coupling appears to be strong in these systems \cite{Steele2018,Ning2019}, which makes a thorough characterization of the nature of the anharmonic lattice vibrations highly desirable also in these materials.  

Generally, the tendency towards octahedral tilting instabilities of the aristotype cubic perovskite structure can be understood in terms of an undersized A site cation, as can be quantified by the Goldschmidt tolerance factor \cite{Goldschmidt1926}. The phase transformations from lower symmetry structures, with tilted octahedra, to the cubic perovskite structure of the single lead halide perovskites has proved to be complex, with tilts of the octahedra away from their high symmetry setting locally present even above the apparent phase transformation temperatures \cite{Bertolotti2017,beecher2016,Cottingham2018}. 

\begin{figure*}[ht!]
    \includegraphics[trim={0.15cm 0.25cm 0cm 0cm},clip,width=2\columnwidth]{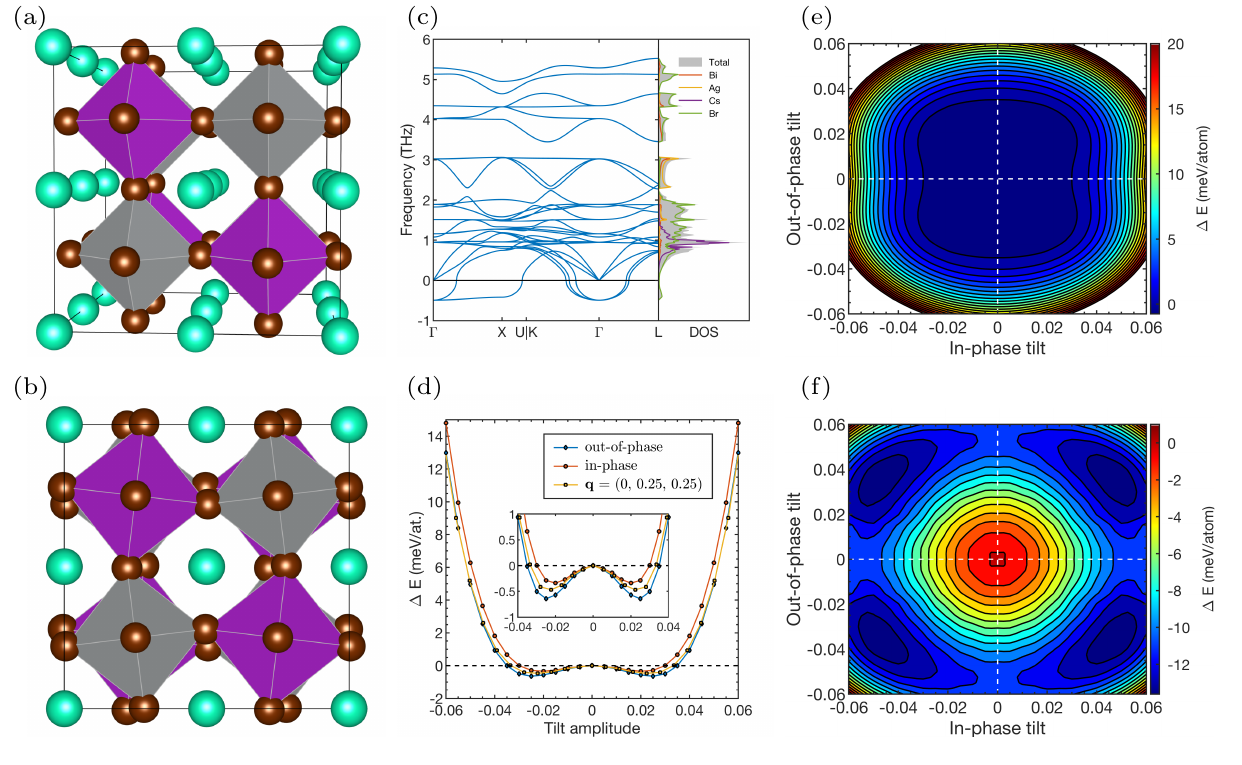}
    \caption{\label{fig:all} (a) Illustration of the cubic double perovskite crystal structure of \cabb. (b) View along the out-of-phase tilting axis of the low temperature tetragonal, a$^-$b$^0$b$^0$, phase. Light green and brown spheres represent Cs and Br ions, respectively, while the Bi and Ag ions are located at the centers of pink and gray octahedra, respectively. (c) Static harmonic phonon dispersion relation and (partial) density of states (DOS) of \cabb. Negative numbers denote imaginary frequencies. (d) 1D potential energy  surface (PES) of three different octahedral tilting modes (see text for details), the inset is a zoom in of the low-energy part of the PES. (e) and (f) 2D a$^-$b$^+$c$^0$ PES of \cabb\ and CsPbBr$_3$, respectively, as a function of the out-of-phase tilt a$^-$ and in-phase tilt b$^+$. The tilt amplitude is given as the offset of one Br ion in units of the lattice constant of the double perovskite (two times the lattice constants of the single perovskite in the CsPbBr$_3$ case).}
\end{figure*}

In the lead-free HDPs, this tendency towards octahedral tilting instabilites remains largely unexplored. While a few HDPs have been shown to be dynamically unstable at 0 K and subsequently stabilized by anharmonic phonon-phonon interactions at room temperature \cite{Zhao2017}, the underlying nature of the anharmonicity and its consequences remain unknown. Furthermore, it was very recently experimentally demonstrated that the prototypical lead free HDP \cabb\ has a cubic-to-tetragonal phase transformation at $T_C \approx $ 122 K upon cooling \cite{Schade2019}, which lacks a theoretical description.

In this Letter, we thoroughly investigate the nature of the lattice vibrations in \cabb\ from first principles. We demonstrate that its lattice dynamics is highly anharmonic even at room temperature, and that the nature of this anharmonicity is qualitatively different from that in the single halide perovskites. The octahedral tilting potential energy surfaces (PESs) of \cabb, in contrast to the single perovskite halide analogous, are very flat, which results in a qualitatively different nature of the dynamics of these octahedra. We fit an effective Hamiltonian from ab initio molecular dynamics (AIMD) and demonstrate that this high anharmonicity necessitates the use of terms up to the 4th order in atomic displacements for an accurate description. We next show how the collapse of a soft zone-center phonon reproduces the experimentally observed phase transformation. Finally, we demonstrate that the softness and anharmonicity of \cabb\ results in an ultra-low lattice thermal conductivity, with a non-conventional $\sim$T$^{-0.5}$ temperature dependence.
%
%

The Density Functional Theory (DFT) calculations were performed in the projector-augmented wave (PAW) \cite{blochl1994} framework as implemented in the Vienna Ab Initio Simulation Package (VASP) \cite{kresse1996,kresse1996_2,kresse1999}. The PBEsol \cite{perdew2008} form of the exchange and correlation functional was used. Static phonon calculations and AIMD simulations were performed using 320 atom supercells while 40 and 80 atom supercells were used for mapping out PESs. The AIMD simulations were done in the NVT ensemble using a Nos\'e-Hoover thermostat with the default Nos\'e mass as set by VASP and a 2 fs timestep. See Supplemental Material \cite{suppl} for full details on computational methodology.      
%
%

We start by investigating the vibrations of \cabb\ in its cubic structure (Fig.\ \ref{fig:all} (a)) under the static harmonic approximation. Fig.\ \ref{fig:all} shows the static harmonic phonon dispersion and density of states (DOS). The most eye-catching feature is the presence of several phonon branches with imaginary frequencies, indicating dynamical instability. In particular, there is an imaginary and essentially dispersionless phonon branch between the $\Gamma$ and $X$ points. The modes on this branch correspond to tilts of the AgBr$_6$ and BiBr$_6$ octahedra. At the X-point the mode is a pure in-phase tilt-pattern i.e., sequential octahedra along the tilt axes are tilted by the same magnitude and in the same direction, a$^+$b$^0$b$^0$ in Glazer notation \cite{Glazer1972}. The $\Gamma$-point mode is a a$^-$b$^-$c$^-$ tilting mode, but with very small b and c axes tilts.

The calculated PESs of in-phase and out-of-phase tilting modes in \cabb\ are shown in Fig.\ \ref{fig:all} (d). As expected, the PESs display minima offset from zero tilt amplitude for both the in-phase and out-of-phase tilting modes. Both minima are, however, very shallow with depths $\sim$0.3 and 0.6 meV/atom for the in-phase and out-of-phase tilting modes, respectively. Thus, even at thermal energies low compared to room temperature, these minima will be easily escaped from and the effective PESs will be very flat. 

At points between $X$ and $\Gamma$ the modes on the imaginary phonon branch correspond to more complicated tilt patterns. At $\mathbf{q} = \left(0,\ 0.25,\ 0.25 \right)$, i.e., halfway between $\Gamma$ and $X$, every second octahedron along the tilt axis stays untilted, while the remaining octahedra are tilted in an out-of-phase fashion (illustrations are given in Fig.\ S4 \cite{suppl}). The PESs for this mode is shown in Fig.\ \ref{fig:all} (d), where we again see a very flat energy landscape. The fact that the different octahedral tilting modes along the soft phonon branch yield very similar, flat, PESs indicates that the bonding between the tilting planes along the tilting axis is weak.

\begin{figure}[ht!]
        \includegraphics[trim={0cm 0cm 0cm 0cm},clip,width=\columnwidth]{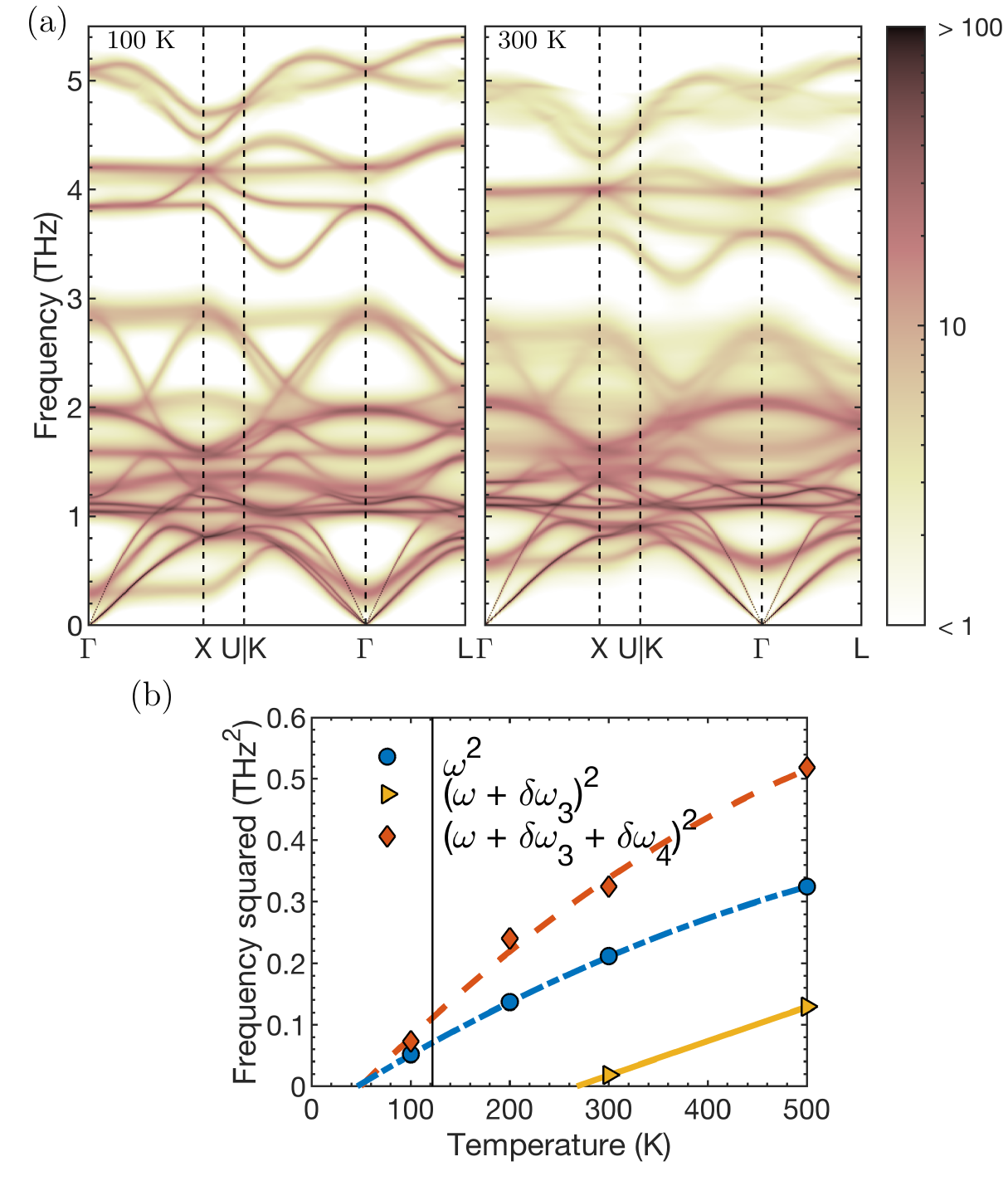}    \caption{\label{fig:sqe} (a) Phonon spectral function, $S \left(\mathbf{q},\Omega \right)$, of \cabb\ at 100 and 300 K. (b) Frequency squared of the soft mode at the $\Gamma$-point with varying corrections, see text for details. The red and blue lines are second order polynomial fits, while the yellow line is merely a straight line to guide to the eye. The vertical black line indicates the experimental transition temperature \cite{Schade2019}}
\end{figure}

Fig.\ \ref{fig:all} (e) shows the 2D a$^-$b$^+$c$^0$ PES, i.e. freezing an out-of-phase tilt around the $a$ axis and in-phase tilt around the $b$-axis, with varying amplitudes. Here, we also optimize the positions of the Cs ions at each fixed tilt configuration. It is instructive to contrast this PES against the same one in CsPbBr$_3$, the lead-containing single-perovskite analog to \cabb, which is shown in Fig.\ \ref{fig:all} (f). There are substantial differences between the PESs in the two systems. While the \cabb\ PES is again very flat around the cubic structure, CsPbBr$_3$ instead has a set of deep minima offset from the cubic structure. This difference can be partly understood from the closer-to-unity Goldschmidt tolerance factor of \cabb\ as compared to CsPbBr$_3$. 

In the single inorganic halide perovskites an emerging interpretation is that the potential energy minima due to octahedral tilting are deep enough that the system occupies them on the timescale of thermal vibrations even above the apparent phase transformation temperature \cite{Bertolotti2017,Bechtel2018,Cottingham2018}. The high symmetry cubic phase then appears as a spatial and/or dynamical average over distinct low symmetry minima. In \cabb, the octahedra instead tilt on flat energy landscapes such that, above the critical temperature, there will be large amplitude Br motion, but the high symmetry phase should be the center of atomic oscillations at any relevant timescale. This implies that the lattice dynamics and the phase transformation in \cabb\, as opposed to e.g.\ CsPbBr$_3$, should be describable in a renormalized phonon picture.

To this end we have performed AIMD simulations for a set of temperatures at equilibrium volumes\cite{suppl} and used the temperature dependent effective potential (TDEP) \cite{Hellman2011,Hellman2013,Hellman2013_2} method to map out an effective fourth-order Hamiltonian, 
\begin{equation} \label{Hamiltonian}
\begin{aligned}
    H &= \overbrace{U_{0} + \sum_{i}\frac{\mathbf{p}_{i}^{2}}{2m_{i}} + \frac{1}{2}\sum_{ij\alpha\beta} \Phi_{ij}^{\alpha \beta}  u_{i}^{\alpha}u_{j}^{\beta}}^\text{$H_{0}$} +\\  &+ \underbrace{\frac{1}{3!}\sum_{ijk\alpha\beta\gamma} \Phi_{ijk}^{\alpha \beta \gamma}  u_{i}^{\alpha} u_{j}^{\beta} u_{k}^{\gamma}}_\text{$H_3$} +  \underbrace{\frac{1}{4!}\sum_{ijkl\alpha\beta\gamma\delta} \Phi_{ijkl}^{\alpha \beta \gamma \delta}  u_{i}^{\alpha} u_{j}^{\beta} u_{k}^{\gamma} u_{l}^{\delta}}_\text{$H_4$}.
\end{aligned}
\end{equation}
Here, $U_{0}$ is a constant energy, $\mathbf{p_{i}}$ and $m_{i}$ are the momentum and mass of atom $i$, respectively, $u_{i}^{\alpha}$ is the displacements of atom $i$, in the cartesian direction $\alpha$, from its position in the reference double perovskite structure and $\Phi_{ij}^{\alpha \beta}$, $\Phi_{ijk}^{\alpha \beta \gamma}$, and $\Phi_{ijkl}^{\alpha \beta \gamma \delta}$ are the elements of the second, third  and fourth order inter-atomic force constants (IFCs), respectively. 

We fit IFCs of increasing order in a sequential fashion, i.e., we first find the best possible fit of the AIMD forces using just the second order IFCs, then fit the third order IFCs to the residual atomic forces, and then similarly for the fourth order IFCs. This fitting scheme ensures that $H_{0}$ in Eq. \ref{Hamiltonian} is the largest, while the effect of $H_3$ and $H_4$ can be treated as perturbations.

From $H_{0}$ we obtain a set of renormalized phonon normal-mode frequencies $\omega_{\mathbf{q}s}$, where $\mathbf{q}$ is the wave vector and $s$ the branch index. To lowest order in perturbation theory $H_3$ and $H_4$ yield, respectively, contributions to the phonon self-energy \cite{Cowley1968} $\Sigma_{\mathbf{q}s}(\Omega) = \Sigma^{(3)}_{\mathbf{q}s}(\Omega) + \Sigma^{(4)}_{\mathbf{q}s} $, where $\Sigma^{(3)}_{\mathbf{q}s}(\Omega) = \Delta^{(3)}_{\mathbf{q}s}(\Omega) + i\Gamma^{(3)}_{\mathbf{q}s}(\Omega)$ is complex and frequency ($\Omega$) dependent, while the contribution from $H_4$, $\Sigma^{(4)}_{\mathbf{q}s} = \Delta^{(4)}_{\mathbf{q}s}$, is purely real and static. The explicit form of this self-energy is well known \cite{Maradudin1962} and is reproduced in our notation in the Supplemental Material \cite{suppl}. Generally, in a "dressed" phonon picture, one might expect the contribution of the fourth order IFCs to be small as their contribution tend to be renormalized into the second order IFCs. Indeed, they are most often not treated or shown to be negligible \cite{Shulumba2017,Errea2019}. As we will show below, however, this contribution turns out to be crucial for \cabb.

From these quantities we obtain a phonon spectral function  $S \left(\mathbf{q},\Omega \right)$\cite{Cowley1968,suppl}, which is shown in Fig.\ \ref{fig:sqe} (a) for temperatures 100 and 300 K. Comparing to the static phonon dispersion in Fig. \ref{fig:all} it is clear that the unstable modes have been anharmonically stabilized. We can also see that at room temperature (300 K) certain phonon branches, in particular in the range $\sim$ 2.2 - 3 THz and the highest two branches between $\sim$ 4.4 - 5 THz, are severely anharmonically broadened. At the $\Gamma$-point these modes are dominated by displacements along the Bi-Br-Ag bonds, indicating a high degree of anharmonicity of these bonds.

Fig.\ \ref{fig:sqe} (b) shows the temperature dependence of the square of the frequency of the soft phonon mode at the $\Gamma$-point. The collapse of this mode should be responsible for the experimentally observed structural transition. Indeed, freezing-in this mode, followed by a full relaxation results in the experimentally observed low temperature \cite{Schade2019} a$^-$b$^0$b$^0$ tilted tetragonal structure (space group I4/m, see Fig.\ \ref{fig:all} (b)), which is $\sim$ 1.0 meV/at.\ lower in energy than the cubic structure.  Low temperature extrapolation of the pure TDEP frequency, i.e. that obtained from $H_{0}$, (blue circles) predicts a phonon collapse at $\sim$ 50 K, which is in fair agreement with the experimentally observed phase transformation temperature of $\sim$ 122 K. We note that the precise frequency of these kind of soft optical modes are known to be sensitive to, in particular, volume and exchange-correlation functional \cite{Errea2019}. 

In Fig.\ \ref{fig:sqe} (b) we also show the temperature dependence of this frequency corrected to lowest order by $H_{3}$ (yellow squares) and by $H_{3}$ and $H_{4}$ (red diamonds) (taken as the positions of the corresponding peak of the spectral function evaluated with the self-energies $\Sigma^{(3)}_{\mathbf{q}s}(\Omega)$ and $\Sigma^{(3)}_{\mathbf{q}s}(\Omega) + \Sigma^{(4)}_{\mathbf{q}s}$, respectively). While the latter case yields a low temperature extrapolation very similar to the pure TDEP frequencies, the former is instead completely different, indicating the importance of the $H_4$ contribution to the effective Hamiltonian in describing \cabb. Further confirmation of this fact can be found by inspecting the low frequency behavior of the acoustic phonon modes, in particular the two transverse acoustic modes in the $\Gamma$-$K$ direction, whose slopes correspond to the elastic constants (in Voigt notation) $C' = (C_{11}-C_{12})/2$ and $C_{44}$, respectively. As we may see from Fig.\ \ref{fig:sqe} these branches are predicted to be essentially degenerate for low frequencies at 300 K, implying elastic isotropy, as can be quantified by a close-to-unity Zener ratio, $C_{44}/C' \approx 1$. We have explicitly confirmed this prediction by calculating the elastic constants from AIMD using deformed supercells and a stress-strain relation \cite{suppl}. Our calculated values are $C_{11} \approx$  29.2 GPa, $C_{12} \approx$ 15.5 GPa and $C_{44} \approx$ 6.9 GPa, giving $A_{Z} \approx 1.01 $. In stark contrast, $S \left(\mathbf{q},\Omega \right)$ evaluated using only $\Sigma^{(3)}_{\mathbf{q}s}(\Omega)$ predicts, incorrectly, substantial elastic anisotropy. Indeed, the later even incorrectly predicts a collapse of one of these transverse acoustic branches around 500 K (See Fig.\ S3 \cite{suppl}).

\begin{figure}
        \includegraphics[trim={0.3cm 0.12cm 0.48cm 0.45cm},clip,width=\columnwidth]{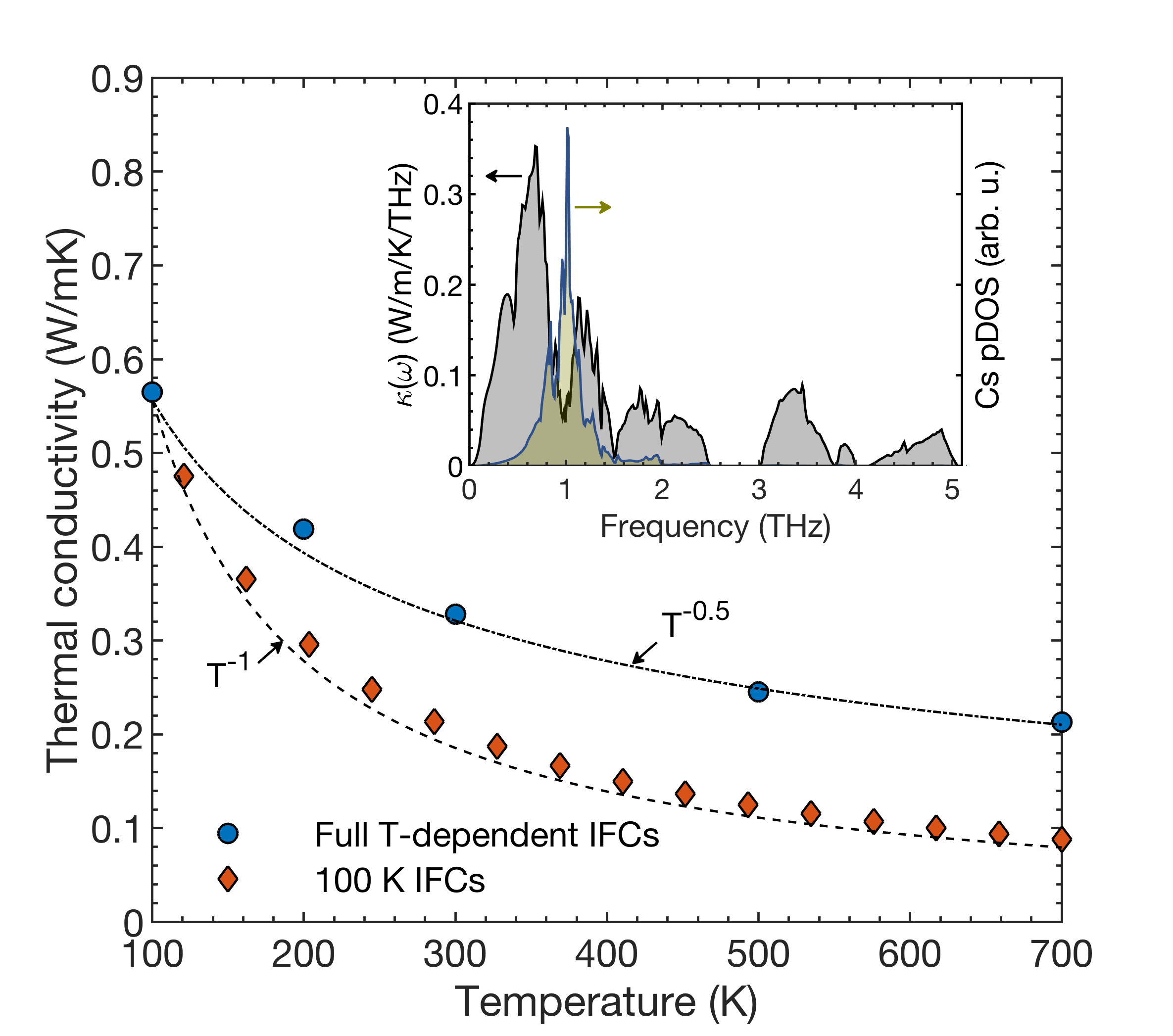}
    \caption{\label{fig:kappa} Temperature dependence of the lattice thermal conductivity of \cabb, evaluated using the full temperature dependent IFCs (blue circles) and using the 100 K IFCs (red squares). The inset shows the spectral thermal conductivity (right axis) and Cs partial DOS (right axis) at 300 K.}
\end{figure}

Having established that \cabb\ is  i) highly anharmonic and ii) very soft (as evidenced by the small elastic constants), we expect a rather low lattice thermal conductivity, $\kappa_l$. We estimate $\kappa_l$\footnote{We note that the thermal conductivity in \cabb\ was recently calculated \cite{Haque2018}, however, incorrectly. The phonon dispersion in Ref. \onlinecite{Haque2018} is totally missing the soft branch.} of \cabb\ by solving the linearized phonon Boltzmann transport equation (BTE) including three-phonon and isotope scattering (details in the Supplemental Material \cite{suppl}). The results are shown by blue circles in Fig.\ \ref{fig:kappa}, where we, as expected, find very low thermal conductivities. Indeed, the 300 K value of 0.33 Wm$^{-1}$K$^{-1}$ firmly establishes \cabb\ as a solid with ultra-low lattice thermal conductivity. Our obtained $\kappa_l$ roughly follows a temperature dependence of T$^{-0.5}$, significantly different from the common T$^{-1}$ dependence usually obeyed by weakly anharmonic solids with thermal conductivities dominated by three-phonon scattering. Such non-standard temperature dependence of thermal conductivities have recently been found in several complex and/or highly anharmonic systems \cite{Romero2015,Xi2019,Wu2019,Shulumba2017} and indicates strong, temperature induced, renormalization of the interacting phonon system responsible for heat transfer. Indeed, we find that a $\sim$T$^{-1}$ behaviour is recovered if we evaluate the thermal conductivity using the 100 K IFCs at all temperatures (red diamonds in Fig.\ \ref{fig:kappa}).  

An additional feature, other than the softness and the anharmonicity, responsible for the low thermal conductivity of \cabb\ can be found by inspecting the spectrally resolved thermal conductivity, $\kappa(\omega)$, shown in the inset of Fig.\ \ref{fig:kappa}. While we can see that, as expected, the majority of the heat is carried by low frequency modes, we also see a marked drop in $\kappa(\omega)$, centered around 1 THz. This drop can be seen to coincide with a sharply peaked Cs partial phonon DOS. In fact, since the Cs ions reside in oversized cubo-octahedral voids in the double perovskite structure, they effectively act as intrinsic rattlers. Correspondingly, the phonon dispersion relation is very flat in a frequency region around 1 THz, as can be appreciated from Fig.\ \ref{fig:all} (c) or Fig.\ \ref{fig:sqe} (a), and we can see the distinctive "avoided crossing" behaviour, characteristic of solids with intrinsic rattlers \cite{Christensen2008,Tadano2015,Lin2016,Jana2017}, for instance of the highest acoustic mode in the $\Gamma$-$L$ direction around 1 THz. 

Ultra-low thermal conductivities have also been demonstrated in both hybrid halide perovskites \cite{Gold-Parker2018} and fully inorganic single halide perovskites \cite{Lee2017}. This has important implications in, for instance, charge carrier cooling. Our results thus firmly put \cabb\ into the set of halide perovskites with ultra-low lattice thermal conductivities. Differently from the single halide perovskites, however, \cabb\ has a high symmetry cubic structure. This combination of a high-symmetry structure (yielding high band-edge degeneracy) and low thermal conductivity is, in fact, quite rare and is believed to be a recipe for high-performance thermoelectric materials \cite{li2018}. 

In summary, we have investigated the anharmonic lattice vibrations of the prototypical lead free halide double perovskite \cabb, in particular in relation to octahedral tilting. We find a branch of unstable phonon modes corresponding to different types of octahedral tilting. Tracing out a set of these modes gives very flat potential energy surfaces. We perform a set of AIMD simulations and fit an effective Hamiltonian, from which we i) reproduce the phonon collapse responsible for the experimentally observed phase transformation, ii) show that \cabb\ is highly anharmonic at room temperature and iii) demonstrate the importance of the high order terms up to fourth order in this Hamiltonian. We finally show that the softness and anharmonic nature of the system result in an ultra-low thermal conductivity, our theoretical estimate is 0.33 Wm$^{-1}$K$^{-1}$ at room temperature.

The support  from the Swedish Research Council (VR) (Project No. 2019-05551) and the Swedish Government Strategic Research Area in Materials Science on Advanced Functional Materials at at Link{\"o}ping University (Faculty Grant SFO-Mat-LiU No.\ 2009-00971) is acknowledged. Theoretical analysis of vibrational properties was supported by the Ministry of Education and Science of the Russian Federation in the framework of MegaGrant No. 074-02-2018-327. The computations were performed on resources provided by the Swedish National Infrastructure for Computing (SNIC) at the PDC Centre for High Performance Computing (PDC-HPC) and the National Supercomputer Center (NSC).

\bibliographystyle{apsrev4-1}
%

\end{document}